\documentclass[11pt]{article}
\usepackage[dvipdfmx]{graphicx}
\usepackage{authblk}
\usepackage{amsmath}
\usepackage{amsthm}
\usepackage{graphics} 
\usepackage{braket}
\usepackage{mathrsfs} 
\usepackage{amsfonts}

\title{
\begin{picture}(0,0)(0,0)%
   \put(330,75){\makebox(0,0)[l]{\textnormal
{\normalsize OU-HET-990
}}}
 \end{picture}
  Why is quantum gravity so difficult (compared to QCD)?}
\author{Hidenori Fukaya\footnote{E-mail address: hfukaya[at]het.phys.sci.osaka-u.ac.jp 
(replacing [at] by @.)}}
\affil{Department of Physics, Osaka University, Toyonaka 560-0043, Japan}
\date{}

\begin{document}
\maketitle
\begin{abstract}
  Gravity is difficult to quantize.
  This is a well-known fact but its reason is given simply by 
  non-renormalizability of the Newton constant and
  little is discussed why among many quantum gauge theories,
  gravity is special.
  In this essay\footnote{
    This essay is an English translation (with some additional discussion)
    of a Japanese article \cite{Fukayaorg} originally
    published in Soryushiron Kenkyu. 
    }, we try to treat the gravity as one of
  many gauge theories, and discuss how it is special and why
  it is difficult to quantize.
\end{abstract}

\section{Introduction}

The author of this article is mainly working on numerical
simulations of lattice QCD.
His ordinary target is a many-body system of quarks and gluons
at energy of a few GeV, and there is no chance for gravity to appear.
But he was asked to teach general relativity to
senior students in a seminar class in 2014
and he needed a lot of study to recover what he all forgot.
In fact, it turned out that no student entered
the high-energy theory lab, which had never happened
in his lab for more than 80 years,
and the class was not opened.

Since the class was not opened,
the author could have stopped his study on gravity and
forget all of them again.
But he took this one hour and half in a week
as a good chance to compare it to QCD
and understand how gravity is different and why it is difficult to quantize.
The difficulty of quantization of gravity is a well-known fact
and it is well explained by the negative mass dimension of
the Newton constant.
However, little is found in the textbooks why only gravity
is special and difficult among many other gauge theories.
As a result of the study in one semester, the author reached
some conclusions and gave a presentation in his lab.
Then his colleague, Kin-ya Oda persuaded him to
write an article and submit it to Soryushiron Kenkyu,
which is a Japanese web journal on high energy particle theory.

General relativity is a gauge theory with
general covariance (and local Lorentz symmetry).
Quantum chromo dynamics (QCD) is a
gauge theory with the color $SU(3)$ invariance.
Both theories are based on the gauge principle,
in which the theory is invariant under
local gauge transformations.
These two theories, therefore, share similar
properties at the classical level.
For example, their Lagrangians are both expressed by
the curvature tensor, which is defined by
the connection gauge field.

There are, however, some differences.
In QCD, the gauge potential or connection $A_\mu$
is the fundamental degrees of freedom (d.o.f.) of the theory,
while in general relativity, the metric $g_{\mu\nu}$ rather
than connection, describes physics.
There is also a big difference in their equation of motion (EOM).
In QCD, the solutions to EOM tend to be a static or
stationary against the time evolution, while in general relativity,
many solutions show a dynamic and strong time dependence,
like inflation, unless we fine-tune the cosmological constant.
In QCD, we can not find  counterpart to vierbein or torsion
in general relativity.
Above all, QCD is known to be renormalizable,
while gravity is, at least, perturbatively non-renormalizable.
Similarities/differences are summarized in Tab.~\ref{tab:similarity}.
\begin{table}[tbh]
  \centering
  \caption{Similarities/differences of QCD and gravity}
  \label{tab:similarity}
  \begin{tabular}{ccc}
    \hline
     & gravity（GR） & QCD\\
    \hline
    connection & $\Gamma_{\mu\nu}^\rho$ & $A_\mu$\\
    curvature & $R^\sigma_{\mu\nu\rho}$ & $F_{\mu\nu}$\\
    fundamental d.o.f. & $g_{\mu\nu}$ & $A_\mu$\\
    Lagrangian & $\sqrt{g}R$ & $\text{Tr}F_{\mu\nu}F^{\mu\nu}$\\
    basic solutions static/stationary? & No. & Yes.\\
    vierbein and torsion? & Yes. & No.\\
    renormalizability & No ! & Yes !\\
    \hline
  \end{tabular}
\end{table}

The author started this study by reading a textbook by
Utiyama \cite{Utiyama}, who was one of great professors
in Osaka university.
In his old work, his first attempt for a unified understanding
of the gravity and Yang-Mills theory is clearly written.
Utiyama also noticed that his attempt is equivalent to
fiber bundle in mathematics, which happened to be developed
almost in parallel at that time.
Then, the author read the textbook by Nash and Sen \cite{NashSen}.
This book indicates that the Riemannian manifold is 
a special case of fiber bundles where the connection is
given by the metric.
In the Kobayashi and Nomizu's textbook \cite{Kobayashi},
it is written why and how the Riemannian manifold is special.
By translating these known facts in mathematics into
physics language, we may be able to explain the difference
between general relativity and other gauge theories.
This essentially corresponds to considering the first
order formalism or Palatini formalism of general relativity \cite{Palatini, Hehl:1994ue, Sardanashvily:2016jhw},
where we use the connection and vierbein as fundamental
fields, rather than the metric. 

In this article, we discuss why and how the special features of
gravity appear in terms of fiber bundles, comparing the
first order formalism of general relativity and other gauge theories.
Then we will reach a conclusion that the difficulty of gravity, including
its non-renormalizability, is caused by a fact:
\[
\mbox{The frame bundle is parallelizable}.
\]
The frame bundle is a kind of a parent of the Riemannian manifold.

The following discussion is, except a few remarks by the author,
not an original consideration but known results in mathematics or physics.
The above conclusion is not an original one either,
but was already given by Heller in Ref.~\cite{Heller}.
The purpose of this article is to draw an attention of the readers
to some non-trivial background behind the explicit fact that
the Newton constant is non-renormalizable.

\section{Fiber bundles}
\label{sec:fiberbundle}

Many textbooks on fiber bundles are available and
even those specially written for physicists exist \cite{NashSen, Nakahara}.
Among them, we recommend the one by Nash \& Sen \cite{NashSen},
which is physicist-friendly, and has nothing written
about the Riemannian manifold,
before the fiber bundle is introduced in its section 7.
Here, we do not rephrase the mathematically precise definition
of the fiber bundles but briefly give a rough sketch of it
using the language of physics, especially of high energy particle theory.

A fiber bundle is a united manifold $E$, which consists of
the base manifold $M$ representing the space-time,
and the fiber space $F$ denoting the space of fields. 
The situation of physics (in 4-dimensions),
where the field is defined on
each point of the space-time, is locally expressed by
a direct product of $\mathbb{R}^4 \times F$.
If this direct product is globally extended,
the total space of the fiber bundle $E$ is
just $M\times F$ and is not very interesting,
but this is not true in general,
and the fiber bundles have non-trivial structures.

In the definition of the fiber bundles,
we have a structure group $G$, which generates the
coordinate transformation in the fiber space $F$.
Although not very stressed in the textbooks,
this $G$ is limited to those linearly acting on the fiber.
This linearity is important as we will discuss later.
For instance, if we take $F$ to be a complex plane,
and $G$ to be $U(1)$ group, then the coordinate transformation
of $\phi(x)\in F$ by $g(x)\in G$ on $x\in M$ is given by
\begin{equation}
\phi'(x)=g(x) \phi(x).
\end{equation}
As clearly seen in this explicit example, the coordinate transformation
in the fiber is just what physicists call the gauge transformation.

We identify a class of fiber bundles which are connected by
the coordinate transformations (both in base and fiber directions).
Its non-trivial structure arises from the consistency condition
among the coordinates (transition function)
on different open patches $U_i$ covering $M$.
The instanton configuration in physics
is a typical example of non-trivial fiber bundles.
If we take the base manifold to be a four-dimensional sphere $M=S^4$,
and a fiber space transformed by $G=SU(2)$,
we need two patches containing north and south poles of $S^4$, respectively.
The overlap region of the two patches can be identified as $S^3$,
and the two ``coordinates'' are related by the gauge transformation,
which maps $S^3 \to G$.
If this map is non-trivial, we cannot express the total fiber bundle
as the direct product $E=M\times F$.

We can take the structure group $G$ itself as a fiber, $F=G$.
In this case, we call this fiber bundle a principal bundle,
often denoted by $P$.
From $P$, we can construct fiber bundles with a fiber $F_G$,
taken in any representation space of $G$,
which is called the associated bundles.
The associated bundle is given by a quotient $E=P\times F_G/G$.
The construction of a principal bundle and its associated bundles,
corresponds to the gauge principle  in physics
that the space-time $M$
and the gauge group $G$ are enough to construct a quantum field theory,
where the fields in various representations
are naturally introduced as the associated vector spaces.

We can give a local structure to a principal bundle by the so-called connection.
The connection of $P$ is given by a decomposition of
the tangent space at a point $u\in P$: $T_u(P)$ into
the one vertical sub-space to the base space (or parallel to the fiber space) $V_u(P)$,
and the other horizontal $H_u(P)$.
This decomposition can be smoothly done at any $u\in P$,
and the total tangent bundle of $P$ becomes $T(P)=V(P)\oplus H(P)$
\footnote{
  The connection is not given by the principal bundle $P$ itself
  but by the tangent bundle $T(P)$, whose base space is $P$.
  The notion of bundles of bundles is difficult to imagine
  for a beginner like the author.
}.

More concretely, the connection of $P$ is obtained by
a differential form known as the connection one-form.
Let us denote the local coordinate on $P$ as
$u=(x,g)$ [$x \in \mathbb{R}^4$, $g\in G$].
Then the connection one form is given by
\begin{eqnarray}
\omega &=& g^{-1}dg + g^{-1}A g,\;\;\;
A = A^a_\mu(x)T_adx^\mu,
\end{eqnarray}
where $A^a_\mu(x)$ is our familiar vector potential,
and $T_a$ are the generators of $G$.
The decomposition of $T(P)$ is achieved by requiring
any vector $X \in H(P)$ to satisfy
\begin{eqnarray}
\label{eq:connection}
\langle \omega, X\rangle =0.
\end{eqnarray}
Here, the degrees of freedom of $\omega$ is the same
as the dimension of $G$.
In a sense, $\omega$ plays a role like a normal vector to $H_u(P)$.

$\omega$ does not change under the coordinate transformation
$g\to hg$ in fiber's direction, 
since we require the gauge field to transform as
\begin{eqnarray}
A \to hdh^{-1} + hAh^{-1}.
\end{eqnarray}
This is nothing but the gauge transformation.
From the connection one-form $\omega$,
we can define the curvature two-form $\Omega$ by
\begin{eqnarray}
\Omega &=& d\omega +\omega \wedge \omega = g^{-1}(dA+A\wedge A)g = g^{-1}Fg,
\end{eqnarray}
where $F$ denotes the field strength.

As a final remark of this section,
we introduce a cheap analogy found on the internet.
Suppose your head as the base manifold $M$
then the fiber $F$ is your hair, and
the total space of the fiber bundle $E$ is your hair style.
We do not know how much nontrivial topological
structures are allowed on your hair bundle.

\section{A fiber bundle view of QCD}

In this section, we discuss how QCD is described by a fiber bundle.
Here we take the base manifold $M$ as a four dimensional flat Euclid space,
which is familiar to the author as a hep-lat person, and
the gauge group $SU(3)$.
These are enough to define the principal bundle $P$ and
introduce its connection.
Here we have not given the metric yet.

How is quantum field theory described with the set-up?
Let us consider a situation where
$P$ with a given connection $A$ appears
according to a kind of probability $\rho$.
This statistical approach,
which is familiar to the author as a hep-lat person,
matches the functional integral in physics.
Here we assume $\rho$ to be a scalar quantity.
The almost unique scalar without using metric is
\begin{eqnarray}
S_\theta = \frac{\theta}{4}\int_M {\rm Tr}F\wedge F,
\end{eqnarray}
which is the $\theta$ term, and it is natural to
assign $\rho = \exp(iS_\theta)$ up to a constant.
It is interesting to note that
$S_\theta$ does not require the metric\footnote{A theory with $\rho = \exp(iS_\theta)$ corresponds to a topological field theory.},
we cannot write down any non-renormalizable action without metric,
and we usually omit $S_\theta$ in QCD, as it is known to be very small.

We should, of course,  assign the metric $g_{\mu\nu}={\rm diag} (1,1,1,1)$
to our base manifold $M$.
This allows us to define the Hodge dual of the curvature two-form,
\begin{eqnarray}
*F_{\mu\nu} &=& \frac{1}{2}F_{\alpha\beta}g^{\alpha \gamma}g^{\beta\delta}\epsilon_{\gamma\delta \mu\nu}.
\end{eqnarray}
Then we can construct the usual gauge action,
\begin{eqnarray}
S_g = \frac{1}{4g^2}\int_M {\rm Tr}F\wedge *F+\cdots,
\end{eqnarray}
where $\cdots$ represents infinitely many types of actions
including non-renormalizable terms.

Now let us introduce the quark field.
As mentioned above, from the principal bundle $P$,
we can construct an associated vector bundle $Q$ as
\begin{eqnarray}
Q = P\times F/G.
\end{eqnarray}
For the quark field, we choose $F$ in the fundamental representation of $SU(3)$.
Using the Dirac operator $D$,  one can define its action
by a gauge invariant scalar as
\begin{eqnarray}
S_q = \int_M d^4 x \sqrt{g}\bar{q} D q.
\end{eqnarray}
It is interesting to note that the (section of) fiber $F$
is reflected in $S_q$, while that of $G$, which is the gauge degrees of freedom,
is absent in $S_g$.

The dynamics of QCD is, thus, described by the statistical mechanics,
where the principal bundle $P$ and the associated bundle $Q$,
which are freely deformed, according to a probability
$\rho = \exp(-S_g +iS_\theta -S_q)$.
The fact that $\rho$ takes an exponential function may be related
to the extensive property of the action or cluster decomposition principles,
but here we do not consider further details.

As a final remark of this section,
we would like to note that lattice gauge theory
have a similar structure to fiber bundle.
It is defined on a discretized lattice space-time,
but the structure group remains the continuum,
and therefore, the fiber space of the principal bundle is continuous.
Assigning the gauge degrees of freedom to the lattice sites
corresponds to giving the fiber space $G$, which determines the principal bundle.
The link variables literally give the ``connection'' between the sites.
By constraining the amplitude of the plaquette,
which smooths the curvature on the lattice,
we can define the instanton number, which cannot be
changed by a link variable variation keeping the smoothness condition on the plaquettes \cite{Luscher:1981zq}.
In fact, even ``subtraction-form'' and its cohomology
can be defined on the lattice \cite{Luscher:1998du},
which plays an important role in a formulation of lattice chiral gauge theories.

\section{Gravity in terms of fiber bundles and solder one-form}

Now let us discuss the (first-order formalism of) general relativity
in terms of fiber bundles.
We consider a four dimensional manifold $M$,
to which we have not given the metric,
and take the real general linear group as the gauge group $G=GL(4,\mathbb{R})$.
This defines a principal bundle called the frame bundle $F(M)$.

The frame bundle has a special property which is not shared by the other general bundles.
It is parallelizable, or equivalently, $T(F(M))$ is always trivial.
A parallelizable manifold is a manifold on which
we can give a globally defined tangent vector space.

For instance, let us consider a two dimensional manifold.
If it is parallelizable, we can globally define the directions
north-south, and east-west.
On a two-dimensional sphere, this is not possible and
we need two singular points: north and south poles.
On a torus, however, it is possible to draw
parallel lines covering the whole torus,
as it is viewed as a flat parallelogram,
whose opposing sides are identified.

The frame bundle on a four-dimensional base manifold $M$ is
a $4+ 4^2=20$-dimensional manifold,
and it is difficult to imagine its parallelizability
but we can show it as follows.

The tangent vector space $T_xM$ at $x\in M$ is $\mathbb{R}^4$
and so does the associated fiber in the fundamental representation
to $GL(4,\mathbb{R})$  principal bundle.
There exists a one-to-one map $e$ between $v\in V$ and $t\in T_xM$
such that
\begin{eqnarray}
\label{eq:solderform}
v^a= e_\mu^a t^\mu.
\end{eqnarray}
The four-component one-form $e=(e_\mu^1 dx^\mu,e_\mu^2 dx^\mu,e_\mu^3 dx^\mu,e_\mu^4 dx^\mu)$
is the so-called solder one-form.
This solder form is nothing but the vierbein in physics.
In the frame bundles, the associated vector space happens to be
isomorphic to the tangent space of the base manifold,
and the solder form is automatically introduced.

To be precise, we should not call $e$ as the solder form.
$e$ is a one-form on $M$ but it is not invariant under coordinate transformations.
The coordinate-independent definition of the solder form $\theta$ is given
as a one-form on $F(M)$ as
\begin{eqnarray}
\label{eq:solderform2}
\theta = g^{-1}e.
\end{eqnarray}
Note that under the gauge transformation $g\to hg$,
$e$ transform as $e\to he$ so that $\theta$ remains invariant.
Here we take the local coordinate of $F(M)$ as $u=(x,g)$\footnote{
Eq.~(\ref{eq:solderform2}) uses a specific coordinate $u=(x,g)$.
A definition without choosing the coordinate basis is given by
requiring $\theta$ for any $X\in T(F(M))$ to satisfy 
\begin{eqnarray}
\langle \theta, X\rangle = \langle e, \pi_*(X)\rangle.
\end{eqnarray}
Here, $\pi_*$ is the induced map $\pi_* :T(F(M))\to T(M)$
of the so-called projection $\pi:F(M)\to M$
(which indicates which point of $F(M)$ corresponds to which point of $M$ whose fiber extends.).
Now $e$ is a pull-back of $\theta$.
This definition looks more difficult than Eq.~(\ref{eq:solderform2})
but it is clearer that $\theta$ is defined on $F(M)$.
}.

The solder form $\theta$ is a one-form on $F(M)$,
but it has non-zero components only in $x$'s directions.
If a tangent vector $X$ at $u$ is in the direction of fiber or $X\in V_u(F(M))$,
we can show $\langle \theta, X\rangle =0$.\footnote{
  In the definition without choosing the coordinate,
  this can be shown from the fact $\pi_*(X)=0$ for $X\in V_u(F(M))$.
}.
As shown in Eq.~(\ref{eq:connection}), the inner-product
between the connection one-form and any vector in $H_u(F(M))$
is zero.
Therefore, for any $X\in T_u(F(M))$, we can conclude that
\begin{eqnarray}
\langle \omega, X\rangle =0\;\mbox{\&}\;\langle \theta, X\rangle=0 
\Longleftrightarrow X=0.
\end{eqnarray}
This means that $X\neq 0$ has non-zero inner product at least,
either with $\omega$ or $\theta$.
Namely, any $X$ can be given by  $\omega$ and $\theta$ as the (dual) basis.
In fact, the degrees of freedom for $\omega$ is $4\times 4=16$
and $\theta$ has 4, and their total 20 matches with the dimension of (the tangent space of) $F(M)$.
Since both of $\omega,\theta$ are smoothly defined on $F(M)$,
any tangent vector field $X$ can be smoothly defined on $F(M)$.
Thus, $F(M)$ is parallelizable.

This situation where we can define not only the connection form
but also the solder form is simply due to
parallelizability of $F(M)$.
This tells us why gravity is special:
the theory of gravity requires not only the gauge field but
also the vierbein as its ingredients.
We also introduce the torsion 2-form,
\begin{eqnarray}
\label{eq:torsion}
\Theta = d\theta + \omega \wedge \theta,
\end{eqnarray}
and it is now clear that the notion of ``torsion''
is special for the frame bundle $F(M)$.

Now our mathematical set-up is ready to make a gravity theory.
Let us start by counting  the degrees of freedom.
The gauge connection field, as an elements of $GL(4,\mathbb{R})$
generators in four different directions, has $4^2\times 4=64$.
The vierbein has $4\times 4=16$.
We have thus 80 in total, which looks quite many compared
to two physical modes of graviton we eventually need.
In the following, we will see how these many degrees of freedom
are dramatically reduced by various physical conditions.

First, we need a reduction of the frame bundle $F(M)$.
The general linear group can be written as
$GL(4,\mathbb{R})=O(4)\times C$, where $C$ is a component
smoothly contractable to a point.
To ignore $C$ and take the reduced structure group $O(4)$
for the principal bundle is called the reduction of the principal bundle.
$O(4)$ corresponds to the (Euclidean version of) local Lorentz group.
This reduction allows us to define a Riemannian metric by
\begin{eqnarray}
g_{\mu\nu}=e_\mu^a e_\nu^b \eta_{ab},\;\;\;\eta_{ab}=\text{diag}(1,1,1,1),
\end{eqnarray}
as $\eta_{ab}$ becomes an invariant tensor under $O(4)$.
We can also introduce the Affine connection
\begin{eqnarray}
\Gamma^\lambda_{\mu\nu}=\left[A^A_\nu\right]^a_b\eta_{ca}e^b_\mu e^c_{\sigma}g^{\sigma\lambda}+(\mbox{differential term}).
\end{eqnarray}
Details of $(\mbox{differential term})$ will be given later.
Here, $A^A_\nu$ is the reduced $O(4)$ gauge field.
It is important to note that $O(4)$ indices $a,b$ are completely
contracted so that $g_{\mu\nu}$ and $\Gamma^\lambda_{\mu\nu}$ are both $O(4)$ invariant.
These objects never appear in QCD, which does not have
vierbein to achieve these contractions.

Here, the general covariance appears as
an emergent gauge symmetry. 
As is well-known, the general covariance preserves
the inner-product of two vector fields,
\begin{eqnarray}
g_{\mu\nu}(x)X^\mu(x) Y^\nu(x),
\end{eqnarray}
under a ``local'' translation, which is achieved by 
a condition on $g_{\mu\nu}$ (metricity condition),
\begin{eqnarray}
\label{eq:mccondition}
\nabla_\rho g_{\mu\nu}\equiv \frac{\partial g_{\mu\nu}}{\partial x_\rho}-g_{\mu\sigma}\Gamma^{\sigma}_{\nu\rho}-
g_{\nu\sigma}\Gamma^{\sigma}_{\mu\rho}=0.
\end{eqnarray}
It is interesting to notice that the general covariance, 
the fundamental property of general relativity appears
as a secondary or emergent invariance of the theory.
It is also interesting that the original $GL(4,\mathbb{R})$ or $O(4)$
gauge invariance is completely hidden unless we consider spinor fields.

One may be confused by counting the degrees of freedom.
After the reduction of the frame bundle,
the $O(4)$ gauge field has $6\times 4=24$,
$g_{\mu\nu}$ has $10$, and that of vierbein which keeps $g_{\mu\nu}$ unchanged
is 6. We have thus 40 in total.
But the metricity condition in Eq.~(\ref{eq:mccondition})
apparently imposes 40 conditions.
If these are independent, the physical degrees of freedom is $80-40-40=0$.
Actually, there should remain 40 here, which means
that the frame bundle reduction must be equivalent to imposing the metricity condition.

The concrete description of
equivalence of the frame bundle reduction and
the metricity condition is hardly found in the literature.
It should be given by some equation which
achieves the fiber bundle reduction and
satisfies the metricity condition at the same time.
In fact, we find that the ``equation of motion(EOM)'' of the vierbein
\begin{eqnarray}
\label{eq:EOMe1}
[D_{\nu} e_{\mu}]^a=(\partial_{\nu} \delta^a_{b} + [A_{\nu}]^a_b) e^b_{\mu}=0,
\end{eqnarray}
is what we require here.
$A_\nu$ is the original $GL(4,\mathbb{R})$ gauge field
and $D_\nu$ denotes its covariant derivative.

Let us decompose the gauge field into its symmetric part
$A_\nu^S$($10\times 4=40$ d.o.f.) and anti-symmetric part
$A^A_\nu$($6\times 4=24$ d.o.f.): $A_\nu=A^S_\nu+A^A_\nu$.
In fact, the condition (\ref{eq:EOMe1}) having 4$^3=$64 non-trivial components
is too strict to constrain only $A^S$.
Therefore, the condition (\ref{eq:EOMe1}) should be understood upto
$O(4)$ gauge transformations to allow the gauge ambiguity with respect to $A^A$.
Then, the above EOM reads
\begin{eqnarray}
\label{eq:EOMe}
(\partial_\nu \delta^a_{b} + [A^A_\nu]^a_b) e^b_{\mu}= -[A^S_\nu+\Delta A^A_\nu]^a_b e_\mu^b,
\end{eqnarray}
where $\Delta A^A_\nu$ represents the $SO(4)$ gauge ambiguity (with 24 d.o.f.).  
This equation having $64-24=40$ non-trivial constraints freezes
$A^S$ as a function of $e_\mu^a$ and $A^A_\nu$ being the connection of the remaining $O(4)$ group,
and achieves the frame bundle reduction from $GL(4,\mathbb{R})$ to $O(4)$ bundles.

Moreover, if we define
\begin{eqnarray}
\Gamma^\rho_{\mu\nu}=-[A^S_\nu+\Delta A^A_\nu]^a_b e^b_\mu [e^{-1}]^\rho_a,
\end{eqnarray}
Eq.~（\ref{eq:EOMe}）is rewritten to an equation
known as the vierbein postulate,
\begin{eqnarray}
\label{eq:vierbeinpostulate}
[\bar{D}_\nu e_{\mu}]^a \equiv (\partial_\nu \delta^a_{b} + [A^A_\nu]^a_b) e^b_{\mu}= \Gamma_{\mu\nu}^\rho e_\rho^a,
\end{eqnarray}
where $\bar{D}_\nu$ the covariant derivative with respect to the $O(4)$ gauge field.
From this postulate, the metricity  condition is automatically obtained,
\begin{eqnarray}
\frac{\partial}{\partial x_\rho}(g_{\mu\nu})= \frac{\partial}{\partial x_\rho}(e_\mu^a e_\nu^b \eta_{ab}) 
=[\bar{D}_\rho e_{\mu}]^a e_\nu^b \eta_{ab} + e_\mu^a[\bar{D}_\rho e_{\nu}]^b \eta_{ab}
= \Gamma_{\mu\rho}^\lambda g_{\lambda\nu}+\Gamma_{\nu\rho}^\lambda g_{\mu\lambda}.
\end{eqnarray}
Note that this metricity condition has no dependence on $\Delta A^A_\nu$, which
appears only in the $\mu$-$\nu$ anti-symmetric part of $\Gamma^\lambda_{\mu\rho}g_{\lambda \nu}$.
Now we have confirmed by Eq.~(\ref{eq:EOMe1}) that the $GL(4,\mathbb{R})$ frame bundle
is reduced to $O(4)$ bundle with the metric which satisfies Eq.~(\ref{eq:mccondition}).
The remaining degrees of freedom is 40.
The Affine connection is given by
\begin{eqnarray}
\label{eq:AffineConnection}
\Gamma^\lambda_{\mu\nu}=[e^{-1}]^\lambda_a  [\bar{D}_\nu e_{\mu}]^a = 
\left[A^A_\nu\right]^a_b\eta_{ca}e^b_\mu e^c_{\sigma}g^{\sigma\lambda}
+(\partial_\nu e_\mu^a)\eta_{ca}e^c_\sigma g^{\sigma\lambda}.
\end{eqnarray}

We need to introduce another important principle of gravity.
It is the equivalence principle, in which we can locally make
the Affine connection zero by the coordinate transformation.
Its necessary and sufficient condition is $\Gamma_{\mu\nu}^\lambda=\Gamma_{\nu\mu}^\lambda$,
or equivalently that the torsion $T^\lambda_{\mu\nu}=\Gamma_{\mu\nu}^\lambda-\Gamma_{\nu\mu}^\lambda$ is zero.
From Eq.~(\ref{eq:vierbeinpostulate}), we can express it by
\begin{eqnarray}
\label{eq:torsionfree}
[\bar{D}_\nu e_{\mu}]^a-[\bar{D}_\mu e_{\nu}]^a =0.
\end{eqnarray}
As it is a two-form, being a pull-back of
the torsion two-form of Eq.~（\ref{eq:torsion}),
the equivalence principle indicates $\Theta =0$, too.
The torsion has 24 degrees of freedom, which happens to be
the same as those of $O(4)$ gauge field.
Hence, we can totally eliminate the $O(4)$ gauge field $A_\mu$ from the theory,
where only the vierbein (16 d.o.f.) is enough to describe it.
The theory looks completely different from the usual Yang-Mills theory.
There is no wonder why gravity is difficult already at this classical level.

Fixing the local Lorentz gauge (6 d.o.f. are lost),
we obtain the conventional form (the second order formalism) of 
general relativity, where we need the metric (10) only.
$\Gamma^\lambda_{\mu\nu}$ becomes the Christoffel symbol,
which is given as a function of the metric.
Further gauge fixing of the general covariance (4)
and four Gauss's law constraints, we can see that the physical
degrees of freedom of gravity is just two.

How can we construct the action to describe the dynamics of gravity?
In the $GL(4,\mathbb{R})$ gauge theory,
we may have the $\theta$ term
\begin{eqnarray}
S_\theta = \frac{\theta}{4}\int_M {\rm Tr}F\wedge F.
\end{eqnarray}
This time, the $GL(4,\mathbb{R})$ group is non-compact
and we do not know if it is still a topological action or not.
It looks that $F=0$ is a solution of the Euler-Lagrange equation of motion.
After the fiber bundle reduction from $GL(4,\mathbb{R})$ to $O(4)$,
and the base manifold $M$ is closed,
this term is related to the Hirzebruch's Signature, which is an integer.

Since we also have the vierbein, we may define a one-form
$\sigma_a^b=(e^{-1})^\mu_a[D_\nu e_\mu]^b dx^\nu$
and construct an action,
\begin{eqnarray}
\int_M \text{Tr}\left[\sigma \wedge \sigma \wedge F\right].
\end{eqnarray}
Although the EOM looks consistent with the least action principle
of this action, it is not found in the literature.

Once the EOM in Eq.~(\ref{eq:EOMe}) is realized by some dynamics
the gauge group is reduced from $GL(4,\mathbb{R})$ to $O(4)$,
and the metric is given, we can write infinitely many actions.
Assuming the conventional dimensional analysis that
the less derivatives in the action indicates the less
ultra-violet divergences, the leading terms would be
the cosmological constant term,
\begin{eqnarray}
\label{eq:cosmological}
S_\Lambda = \Lambda M_\text{pl}^2 \int_M e^a \wedge e^b \wedge e^c \wedge e^d \epsilon_{abcd},
\end{eqnarray}
and the next-to-leading one is the Einstein-Hilbert action,
\begin{eqnarray}
\label{eq:EH}
S_{EH} = M_\text{pl}^2 \int_M e^a \wedge e^b \wedge [\bar{D}A^A]^c_d \eta^{de}\epsilon_{abce},
\end{eqnarray}
where $M_\text{pl}$ is the Planck scale.
Note again that existence of these curvature independent and
linear terms is special for gravity.
Compared to QCD, where the terms in even products of $\bar{D}A^A$ are allowed,
the gravity action looks unstable, which explains
why many non-stationary solutions exist for the Einstein equation.
Also, we can construct the matter field action such as 
\begin{eqnarray}
S_m = \int_M d^4x \bar{\psi}g^{\mu\nu}\gamma_a e_\mu^a (\partial_\nu +[A^A_\nu]^{b}_c\eta^{cd}\gamma_b \gamma_d )\psi(x),
\end{eqnarray}
where $\gamma_a$ are $4\times 4$ Dirac matrices.

Let us assume that the other possible higher derivative terms
are all negligible, and consider the least action principle
of $S=S_\Lambda + S_{EH} + S_m$.
From the variation of $A^A_\mu$
(where we further assume that the fermion field does not have
non-scalar condensate),
the torsionless condition (\ref{eq:torsionfree}) is derived.
In a sense, the equivalence principle is not needed to be given
a priori, but can be realized by the dynamics of the Affine connection.
The variation of $e^a_\mu$ leads to the conventional Einstein equation
or the second order formalism of the general relativity\footnote{
  From Eq.~(\ref{eq:AffineConnection}), one obtains
  the Riemannian tensor as
$R^\lambda_{\mu\nu\rho}=[e^{-1}]^\lambda_a [\bar{D}_\rho A_\nu^A]^a_b e_\mu^b$.
}.
We summarize the above discussion in Tab~\ref{tab:1st2nd}.

\begin{table}[tbh]
  \centering
  \caption{How to obtain the conventional general relativity
    from the 1st order formalism. (g.c. denotes general covariance)}
  \label{tab:1st2nd}
  \begin{tabular}{p{3cm}cccc}
    \hline
　conditions &　equations    & gauge sym. & fields　　　& d.o.f.\\
    \hline
  frame bundles &---& $GL(4,\mathbb{R})$ & $[A_\mu]^a_b, [e_\mu]^a$ & 80\\ $\downarrow$\\
  \shortstack{bundle reduction\\(metricity)} & $[D_{\nu} e_{\mu}]^a=0$ & $O(4)+$ g.c. &  
\shortstack{$[A^A_\mu]^a_b, [e_\mu]^a$\\ ($g_{\mu\nu}$)} & 40\\ $\downarrow$\\
  \shortstack{equivalence principle\\(zero torsion)} & $[\bar{D}_\nu e_{\mu}]^a-[\bar{D}_\mu e_{\nu}]^a =0$ & $O(4)+$ g.c. & \shortstack{$[e_\mu]^a$\\($g_{\mu\nu}$)} & 16 \\ $\downarrow$\\
$O(4)$ gauge fixing & many ways & g.c. &$g_{\mu\nu}$ & 10 \\
    \hline
  \end{tabular}
\end{table}

\section{Frame bundle reduction and Higgs mechanism}

In this section\footnote{This section is new and not contained in the original Japanese version.}, we discuss that the fiber bundle reduction
can be in general viewed as the Higgs mechanism.
In gravity theory at our hand, if we identify the vierbein as
a Higgs field \cite{Sardanashvily:2016jhw}, the EOM (\ref{eq:EOMe}) is naturally obtained.

Let us start with summarizing the argument in
Kobayashi and Nomizu's textbook \cite{Kobayashi}
on how the principal bundle is reduced.
It is given in the following two steps.
\begin{enumerate}
\item A principle bundle $P(G, M)$ (where $G$ is its structure group and $M$ is the base manifold) is reducible to $P(H, M)$ where $H$ is a sub-group of
  $G$, if and only if the associated bundle $E(G/H, M, G)$ admits a section.
\item The connection of $P(H,M)$ is uniquely given by
  the condition of the section of $E(G/H, M, G)$ to be parallel to
  the connection of $P(G,M)$.
\end{enumerate}
Here $E(G/H, M, G)$ is a fiber bundle associated to the original
principal bundle, whose fiber space is the quotient $G/H$.

In fact, this principal bundle reduction is nothing
but the Higgs mechanism in physics.
The first step is interpreted as a Higgs field taking
a vacuum expectation value (VEV) in $G/H$.
We know in this case, the original $G$ gauge symmetry
is broken down to $H$. The second condition determines
which part of the original gauge field remains as
the $H$ gauge field.

Following \cite{Honda:1979py}, let us confirm the above argument
taking a simplest example, $G=SO(3)$ gauge theory with a Higgs,
or equivalently, $SU(2)$ gauge theory with an adjoint Higgs field.
In the standard gauge, we take a constant Higgs VEV
$h_0 = (0,0,v)^T$ but if we allow its gauge transformation,
we can have a $x$-dependent VEV
\begin{equation}
h(x) = g(x)h_0, 
\end{equation}
which, in fact, determines a section of $G/H$ bundle,
since $g(x)\in SO(2)$ keeps $h_0$ invariant.
Here, the first step of Kobayashi and Nomizu is done.

Next let us identify the remaining $SO(2)$ gauge field.
The second step of Kobayashi and Nomizu tells us that
the section of $E(G/H, G, M)$ or the Higgs field
should be parallel to the original connection of $G$ gauge theory.
Namely, the condition is simply given as the Higgs ``EOM''\footnote{
  This condition is stronger than the usual EOM
  $g^{\mu\nu}D_\mu^{SO(3)}D_\nu^{SO(3)}h(x) = 0$.
  },
\begin{equation}
  \label{eq:SO3EOM}
D_\mu^{SO(3)}h(x) = 0,
\end{equation}
where $D_\mu^{SO(3)}$ is the covariant derivative
with respect to the original $SO(3)$ gauge field.
Under this condition, two components of the original
$SO(3)$ gauge fields are frozen, becoming functions of
$h$ and the remaining $SO(2)$ (or $U(1)$) gauge field,
which is given by
\begin{equation}
A_\mu^{SO(2)} = \frac{\hat{h}^T \left[ A_\mu + \hat{h} \times \partial_\mu \hat{h}\right]}{\hat{h}^T \hat{h}},
\end{equation}
where $A_\mu$ is the original $SO(3)$ gauge field,
and $\hat{h}=h/v$ is dimensionless expression of the Higgs.

For the simplest case $\hat{h}(x)=(0,0,1)^T$,
$A_\mu^{SO(2)}$ is just the third component of $A_\mu$, while
other two components become zero
(W bosons are not excited in classical theory).
If the Higgs field has a singular point, $h(x_0)=0$,
then on a two-dimensional sphere $S^2$ around $x_0$ makes
a map: $S^2\to SO(3)/SO(2)=S^2$, labeled by an integer,
which gives the magnetic charge of the 'tHooft-Polyakov monopole.

In this way, the reduction of the principal bundles
can be interpreted as the Higgs mechanism in physics.
Now let us get back to the gravity theory and discuss
how the frame bundle reduction is realized in terms of the Higgs mechanism.

One immediately finds that the Higgs EOM Eq.~(\ref{eq:SO3EOM})
in $SO(3)$ theory looks like that of the vierbein in Eq.~(\ref{eq:EOMe}).
Moreover, the vierbein can be written as $e_\mu^a(x) = [g'(x) \bar{e}_\mu]^a$,
where $\bar{e}$ is a constant fixed back-ground,
and $g'(x)$ is the $GL(4,\mathbb{R})$ gauge transformations,
and the metric 
\begin{eqnarray}
  g_{\mu\nu} =  [g'\bar{e}_\mu]^a[g'\bar{e}_\nu]^b  \eta_{ab},
\end{eqnarray}
determines a section of $E(G/H, G, M)$ bundle
where $G=GL(4,\mathbb{R})$ and $H=O(4)$.
Note that if $g'\in O(4)$, the metric is unchanged.
Namely, it is the vierbein that reduces the frame bundle
from $GL(4,\mathbb{R})$ to $O(4)$ via the Higgs mechanism,
and the metric is the Higgs VEV\footnote{
  The idea of metric as the Higgs VEV is
  not new but found in the literature.
 }. Eq.~(\ref{eq:EOMe}) is the correct equation for achieving this.

In the conventional Higgs mechanism from $G\to H$,
we can easily construct a $G$-invariant Higgs action,
which spontaneously produces a VEV of the Higgs field.
But for gravity, as discussed in the previous section,
it looks difficult to make a $G$-invariant action,
since we cannot use the metric, which breaks
the original $G=GL(4,\mathbb{R})$ symmetry.
The non-compactness of $GL(4,\mathbb{R})$ may also be
a problem in constructing a quantum theory.

\section{Why is quantum gravity difficult?}

Finally let us consider why quantization of gravity is difficult.
As seen in the previous sections, gravity theory differs
from other gauge theories very much already at classical level.
There is thus little doubt that quantization
of gravity is very difficult, too.
The difficulty of gravity at classical level is due to the
fact that the frame bundle is parallelizable, introducing
the vierbein in the theory.
Here, we see that the vierbein is also a obstacle for quantization.
The focus of this article is to shed light on the difficulty
of gravity, thus we do not review many previous attempts to quantize gravity.

We start with agreeing Nakanishi's claim (in \cite{Nakanishi})
that it is not appropriate to quantize the metric $g_{\mu\nu}$.
As seen in the previous sections, from the fiber bundle picture,
the metric is not a fundamental field but a composite
of the vierbein.
It is analogous to the pions in QCD.
The effective theory of pions is known as chiral perturbation theory,
which is not renormalizable.
But we never worry about the renormalizability of QCD itself.

Clearly, the vierbein is a vector field, having spin 1.
It is almost obvious that spin 1 particle is easier to treat
than higher spin particles.
In our textbooks, we usually treat spin 0,1/2 and 1, only \cite{Srednicki}.
The cosmological constant term in Eq.~(\ref{eq:cosmological}) and
the Einstein-Hilbert action in Eq.~(\ref{eq:EH}) in terms of vierbein
look different from the conventional action with metric, which
looks hopelessly non-renormalizable.
Since there is, apparently at least, no coefficient
with negative mass dimensions, we may consider the former as
four-point self interactions of vierbein, and the latter as
the 3-point interaction with the gauge field.
One may feel like that if we introduce an appropriate
kinetic term of vierbein and gauge connections,
we may construct a renormalizable quantum field theory of them.

However, our life is not so easy.
The above situation is analogous to the one where
we add a charged vector field (like $\rho$ meson) to QED.
It is well-known that we cannot construct a renormalizable
field theory with the charged vector field, as
its longitudinal modes produces a UV divergence.
This divergence can be removed only when the additional
charged vector fields are the gauge fields of another gauge symmetry
(like the weak bosons of $SU(2)$ theory,
whose mass is given by the Higgs mechanism).
This observation suggests that we cannot quantize gravity
unless we find another gauge symmetry, in which
the vierbein becomes its gauge connection.

In fact, there are many attempts to formulate a theory containing
the vierbein as a gauge field in the literature.
A successful example is the three-dimensional gravity.
Witten showed that it is renormalizable \cite{Witten:1988hc},
where the dreibein plays a role of gauge bosons.
In three-dimensional gravity, we can treat
the dreibein as a generator of local translation,
which differs from general coordinate transformation,
and the action has a form of a Chern-Simons term, which is invariant
under the new translation symmetry.
There happens to be an extended gauge symmetry,
where the dreibein is treated as its gauge field,
and we can make the theory renormalizable.

However, this is not the case in four or higher dimensional gravity.
The Einstein-Hilbert type actions cannot be invariant
under this local translation.
There is a fundamental difficulty in mathematics, too.
As mentioned in Sec.~\ref{sec:fiberbundle},
the structure group is required to act linearly on the fibers.
The local translation, acting non-linearly on
fibers, cannot be a structure group.
We need an extended mathematical set-up beyond fiber bundles.

It is also non-trivial to figure out
which steps in Tab.~\ref{tab:1st2nd} are
given as fundamental principles of the theory,
and which are given by dynamics,
during the reduction from the first order formalism
to the second order.
The lower parts, like torsion less condition,
look easer to be embedded as
dynamical consequences in the theory,
while it is difficult to imagine how the upper parts,
especially, the Higgs mechanism is realized.

\section{Conclusion}

The difficulty of gravity lies in the fact that
its basic mathematical object, frame bundle, is parallelizable.
This parallelizability introduces the vierbein field,
in addition to the conventional gauge field.
The vierbein introduces unfamiliar physical observables,
such as torsion, Affine connections, and
make different types of the gauge invariant actions.
In particular, the Einstein-Hilbert action,
which is linear in the curvature, originates from
the contraction of the two vierbein fields
and a curvature field.

The vierbein is a spin 1 particle,
which cannot be treated as a gauge particle.
This explains why the renormalization is difficult.
An exception is the three-dimensional gravity,
where dreibein can be treated as a gauge field of
local translation. 
However, it looks just a lucky coincidence where
the action happens to be a Chern-Simons term.
In general, there is a difficulty in mathematics:
the local translation cannot be
incorporated as a structure group of fiber bundles.

These are our conclusions.
As being a non-expert, the author's discussion may be too naive and simple.
It may be totally wrong to start with the frame bundle,
as a target to quantize the gravity.
Or our arguments above may be trivial for some experts.
There may be more fundamental problems in this article.
However, the author is, at least, satisfied by understanding
that QCD would never have something corresponding to the vierbein.

The author thanks Kin-ya Oda for suggesting
to write this article,  Shigeki Sugimoto for
discussion about the physical meaning of vierbein,
Norihiro Tanahashi for checking this article
from a view by a gravity expert,
Akinori Tanaka for teaching why three-dimensional sphere
is parallelizable, and Satoshi Yamaguchi
for helping the author in reading mathematical textbooks.

\end{document}